\documentclass[aps,reprint,prb,longbibliography,superscriptaddress]{revtex4-1} 
\usepackage{amsmath}
\usepackage{amsfonts}
\usepackage{latexsym}
\usepackage{graphicx}
\usepackage{color}
\usepackage{bm}
\usepackage{float}

\definecolor{Blue}{rgb}{0,0,1}

\newcommand{\ket}[1]{{\left\vert{#1}\right\rangle}}

\usepackage{hyperref} 
\usepackage[all]{hypcap}
\hypersetup{pdfauthor = {First Author},pdftitle = {Experimental Test of Phase Commutativity}}

\begin{document}
\title{Experimental Test of Hyper-Complex Quantum Theories\\
}

\author{Lorenzo M. Procopio$^\ddag$}
\affiliation{Faculty of Physics, University of Vienna, Boltzmanngasse 5, A-1090 Vienna, Austria}

\author{Lee A. Rozema$^\ddag$}
\affiliation{Faculty of Physics, University of Vienna, Boltzmanngasse 5, A-1090 Vienna, Austria}

\author{Zi Jing Wong$^\ddag$}
\affiliation{Nanoscale Science and Engineering Center, University of California, Berkeley CA, USA}
\affiliation{Materials Sciences Division, Lawrence Berkeley National Laboratory, Berkeley CA, USA}

\author{Deny R. Hamel}
\affiliation{Faculty of Physics, University of Vienna, Boltzmanngasse 5, A-1090 Vienna, Austria}
\affiliation{D\'{e}partement de physique et d'astronomie, Universit\'{e} de Moncton, Moncton, New Brunswick E1A 3E9, Canada}

\author{Kevin O'Brien}
\affiliation{Nanoscale Science and Engineering Center, University of California, Berkeley CA, USA}
\affiliation{Materials Sciences Division, Lawrence Berkeley National Laboratory, Berkeley CA, USA}

\author{Xiang Zhang}
\thanks{xiang@berkeley.edu}
\affiliation{Nanoscale Science and Engineering Center, University of California, Berkeley CA, USA}
\affiliation{Materials Sciences Division, Lawrence Berkeley National Laboratory, Berkeley CA, USA}

\author{Borivoje Daki\'c}
\affiliation{Faculty of Physics, University of Vienna, Boltzmanngasse 5, A-1090 Vienna, Austria}
\affiliation{Institute for Quantum Optics and Quantum Information, Austrian Academy of Sciences, Boltzmanngasse 3, A-1090 Vienna, Austria\\
$^\ddag$These authors contributed equally to the work.}

\author{Philip Walther}
\thanks{philip.walther@univie.ac.at}
\affiliation{Faculty of Physics, University of Vienna, Boltzmanngasse 5, A-1090 Vienna, Austria}


\begin{abstract}
In standard quantum mechanics, complex numbers are used to describe the wavefunction.
Although complex numbers have proven sufficient to predict the results of existing experiments, there is no apparent theoretical reason to choose them over real numbers or generalizations of complex numbers, i.e. hyper-complex numbers.
Experiments performed to date have proven that real numbers are insufficient, but whether or not hyper-complex numbers are required remains an open question.
Quantum theories based on hyper-complex numbers are one example of a post-quantum theory, which must be put on a firm experimental foundation.
Here we experimentally probe hyper-complex quantum theories, by studying one of their deviations from complex quantum theory: the non-commutativity of phases.
We do so by passing single photons through a Sagnac interferometer containing two physically different phases, having refractive indices of opposite sign.
By showing that the phases commute with high precision, we place limits on a particular prediction of hyper-complex quantum theories.
\end{abstract}

\maketitle

\section*{Introduction}

Quantum mechanics is an extremely well-established scientific theory. 
It has been successfully tested against competing ``classical'' theories for almost 100 years.
Such classical-like theories attempt to maintain elements of classicality;
examples include hidden-variable theories \cite{Einstein1935,bell1964,weihs1998,Christensen2013,giustina2013,hensen2015loophole,giustina2015significant,shalm2015strong}, non-linear extensions of quantum mechanics\cite{Bialynicki1976,Weinberg1989}, and collapse models\cite{grw}.
Recently, ``post-quantum theories'', which still possess inherently  quantum features, such as superpostion and ``non-locality'', have been proposed\cite{Sorkin1994,dakic2014density,finkelstein1962,adler1995quaternionic,Horwitz1996,Kanatchikov1999,baez2012division}.
Experimental tests of these post-quantum theories have only recently begun \cite{Kaiser1984,klein1988,Sinha2010}.
In one class of post-quantum theories---so-called hyper-complex quantum theories \cite{finkelstein1962,adler1995quaternionic,Horwitz1996,Kanatchikov1999,garner2014quantum,baez2012division}---simple phases do not necessarily commute\cite{Peres1979,Kaiser1984}.
Here, we experimentally search for this effect by applying two phases to single photons in a Sagnac interferometer.
We induce the two phases by very different optical media to enhance any potential non-commutativity.
One phase is a standard optical phase induced with a liquid-crystal, and the other is a negative phase which is induced by an artificial nanostructured  metamaterial \cite{dolling2007negative,shalaev2007,yao2008,Zhang2011}.
We find a null result, meaning that the net phase when applying the two phases in either order (meta-material before liquid crystal or vice versa) is equivalent to within at least $0.03^\circ$.
This bound is one order of magnitude tighter than previous experimental work\cite{Kaiser1984}.
Our experiment demonstrates the combination of a broadband, negative-index metamaterial with single-photon technology at optical wavelengths.
Furthermore, we place bounds on the non-commutativity of phases within any hyper-complex quantum theory.

The superposition principle states that linear combinations of wavefunctions are also valid wavefunctions.
In textbook quantum mechanics these weighting coefficients are complex numbers, but there is no immediate theoretical requirement for this restriction.
{For example, it was shown by Birkhoff and von Neumann in 1936 that a mathematically consistent quantum theory can be constructed using only real numbers \cite{Birkhoff1936}, but such a theory cannot correctly predict the results of certain experiments. 
One well-known example of this failure is that complex numbers are required to model all physically-realizable two-level systems, such as the polarization state of a photon.} 
So far ``complex quantum mechanics'' (CQM) has proven necessary to describe most quantum phenomena, but it is not known if it will remain sufficient.

Similarly, one can construct a quantum theory based on hyper-complex numbers \cite{finkelstein1962,adler1995quaternionic}, such as quaternions \cite{hamilton1844}.
A quaternion is a mathematical generalization of the complex number with three, rather than one, imaginary components.
Quaternionic quantum mechanics (QQM) has attracted much attention, in part because {it is a natural and elegant extension of standard quantum theory} \cite{klein1988,finkelstein1962,adler1995quaternionic,Horwitz1996,Kanatchikov1999,Peres1979,Kaiser1984,brumby1996, baez2012division}.
{Unlike other post-quantum theories, QQM does not necessarily modify the postulates of quantum mechanics\cite{Sorkin1994,popescu1994quantum,paterek2010theories,barrett2007information}.}
However, QQM makes certain experimental predictions which are different from predictions of complex quantum mechanics, just like the predictions of a real quantum theory disagree with those of a complex theory.

One disagreement between CQM and QQM is the (non) commutativity of phases.
In CQM  phases commute, since they are described by complex numbers.
However, quaternions do not commute in general, thus in QQM phases will not necessarily commute.
Based on this idea, in 1979 Asher Peres proposed several experimental tests to search for quaternions in quantum mechanics\cite{Peres1979}.
Because of technological limitations at the time, only a single neutron experiment has tested his ideas \cite{Kaiser1984}.
Inspired by Peres, here we present an experiment that allows us to precisely search for the phase non-commutativity predicted by QQM. 
To do so we exploit modern photonic quantum technologies, which provide a proven platform for foundational tests\cite{weihs1998,Christensen2013,giustina2013,hensen2015loophole,giustina2015significant,shalm2015strong,kocsis2011, rozema2012, Shadbolt2014}.

\begin{figure}
\begin{center} 
\includegraphics[width = \columnwidth]{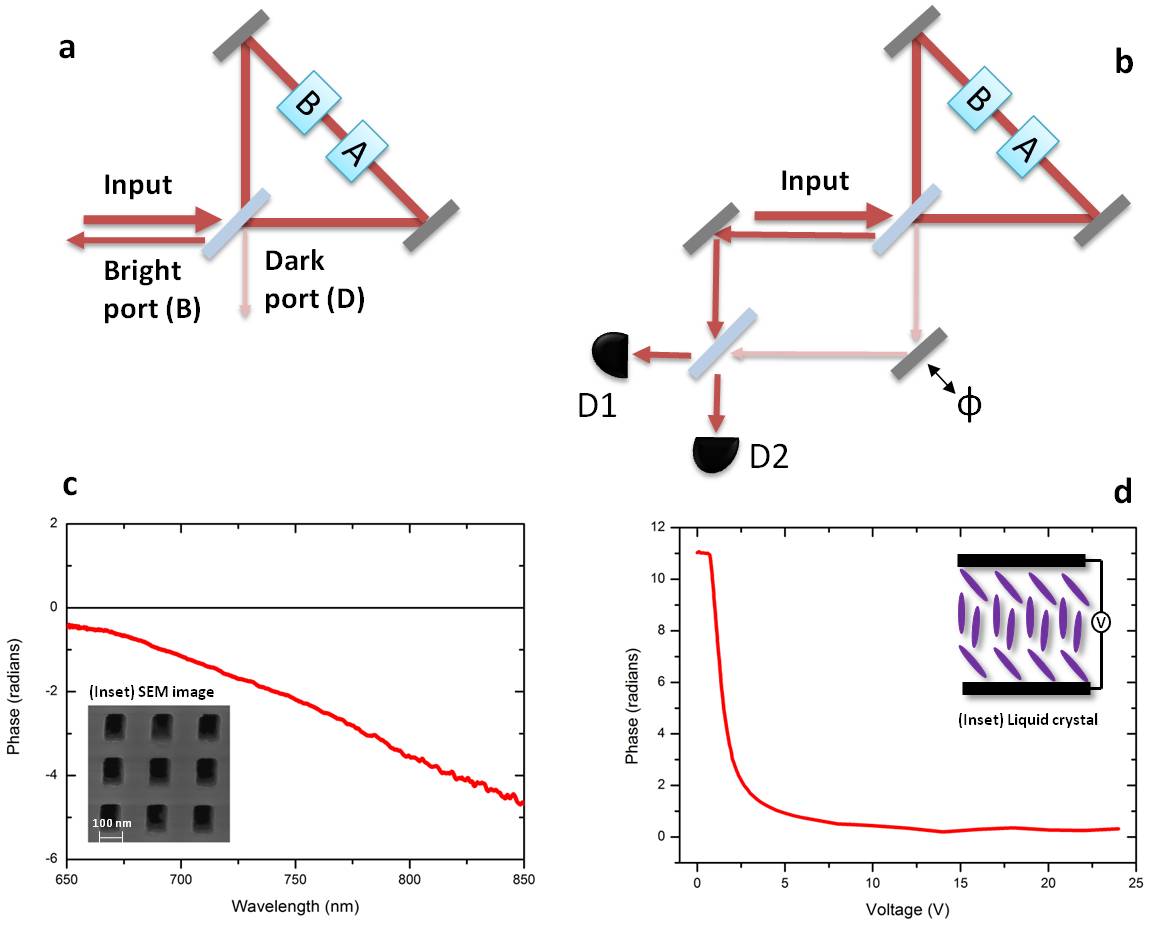}
\label{cartoon}
\end{center}
\caption{\textbf{Experimental schematic and phase characterization.} (\textbf{a}) If two different phases $A$ and $B$ are placed inside a Sagnac interferometer and, if the phases commute, all the incoming light should exit through the ``bright port'', while there should be no light in the ``dark port''. If $A$ and $B$ do not commute the dark port will not be dark. (\textbf{b}) Adding a Mach-Zender interferometer to interfere the bright and dark ports allows for a more precise measurement of the leakage into the dark port. (\textbf{c}) Wavelength dependence of the phase shift of our negative index metamaterial. 
For the wavelength of our single photons, 790 nm, the measured phase is about -$\pi$, which corresponds to a refractive index of the multilayer fishnet of -0.4. Inset: SEM image of the negative index metamaterial. (\textbf{d}) Phase response of the nematic liquid crystal.  The measured relative phase (modulo 2$\pi$) between the LC and the air for transmitted light is about $+\pi$. Inset: representation of a liquid crystal.}
\end{figure}

\section*{Experimental Proposal}

Our experiment is based on a Sagnac interferometer containing different phases.
As  illustrated in Fig. \ref{cartoon}a, a perfectly balanced Sagnac interferometer, with an even number of reflections, results in all of the photons exiting through the same port that they entered.
This results in a ``bright port'' and a ``dark port''.
However, this assumes that the phases commute, as CQM dictates.
To be more specific, let $A$ and $B$ be two phase operators $A = \alpha\mathcal{I}$, and $B = \beta\mathcal{I}$ (where $\mathcal{I}$ is the identity operator).
In CQM $\alpha$ and $\beta$ are complex numbers, but in general they could be quaternions, or other hyper-complex numbers.
Then the probability to detect a photon in the dark port, in an ideal interferometer with no experimental imperfections, depends on the commutator of $\alpha$ and $\beta$ as
\begin{equation}\label{eq:PdarkID}
P_\mathrm{D}^\mathrm{ideal}= \frac{|[\alpha,\beta]|^2}{4}.
\end{equation}
\noindent  In CQM, $\alpha=e^{i\phi_A}$ and $\beta=e^{i\phi_B}$ are complex numbers, where $\phi_A$ and $\phi_B$ are real numbers.
In this case $P_\mathrm{D}^\mathrm{ideal}=0$.
On the other hand, in QQM $\alpha$ and $\beta$ are quaternions, which do not generally commute; hence, we expect that $P_\mathrm{D}^\mathrm{ideal}$ can deviate from $0$. 
See the Appendix, Eq. \ref{eq:almostComm} for more details.

\begin{figure*}
\begin{center} 
\includegraphics[width = \textwidth]{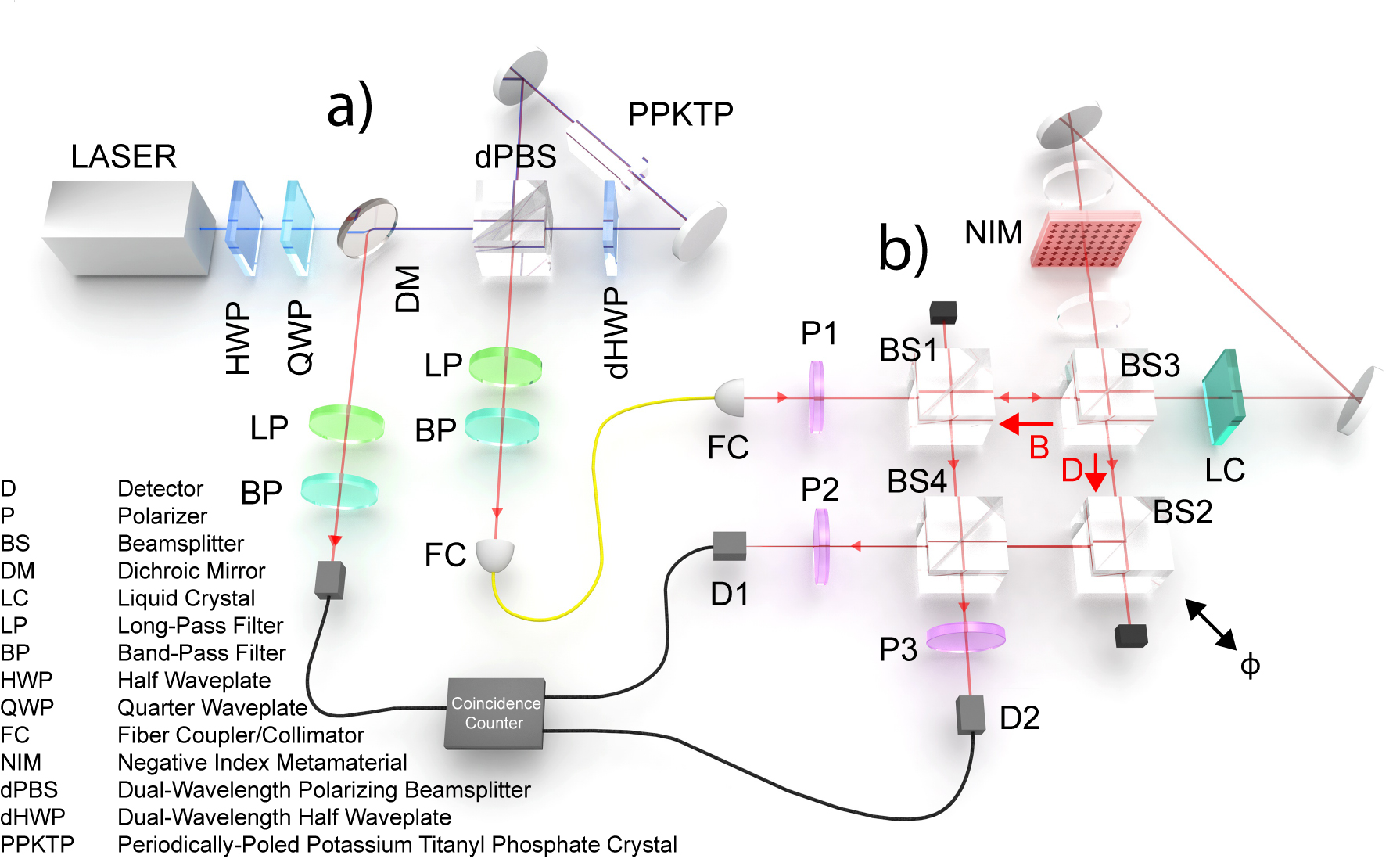} 
\label{expt}
\end{center}
\caption{\textbf{Experimental apparatus}. A detailed schematic of experiment to search for a quaternionic contribution to phase shifts. (\textbf{a}) We generate photon pairs in a separable polarization state. One photon is used to herald while the other one is sent to our interferometers. (\textbf{b}) We couple a Sagnac interferometer into a Mach-Zehnder interferometer to search for non-commuting phases. We monitor the interference in the Mach-Zehnder interferometer as its phase $\phi$ is scanned. The output photons are detected using single-photon detectors D1 and D2.  The detectors are connected to coincidence logic to herald single photons.  Two phases are applied inside the Sagnac which we can controllably ``turn on'' and ``turn off''.  The liquid crystal (LC) is controlled by applying voltage to it, and the negative index metamaterial (NIM) is mounted on a motorized translation stage so it can be ``turned off'' by physically removing it from the interferometer.}
\end{figure*}

In practice, photons can also leak into the dark port because of experimental imperfections.
We can quantify the imperfections of the Sagnac interferometer by a visibility, defined as $v=(P_\mathrm{B}-P_\mathrm{D})/(P_\mathrm{B}+P_\mathrm{D})$, that is less than $1$.
Here, $P_\mathrm{D}$ ($P_\mathrm{B}$) is the probability to detect a photon in the dark (bright) port.
In the Appendix we show that, for such an imperfect Sagnac interferometer with two non-commuting phases, $P_\mathrm{D}$ is $P_\mathrm{D}=\frac{1}{2}-\frac{v\Gamma}{2}$, where 
\begin{equation}\label{eq:PIdark}
\Gamma=1-\frac{|[\alpha,\beta]|^2}{2}.
\end{equation}

Since we expect any deviation from CQM to be small, we expect $P_\mathrm{D}$ to be small.
Thus, we measure an amplified signal by interfering the bright and dark ports of the Sagnac interferometer in a Mach-Zehnder--like interferometer (Fig. \ref{cartoon}b).
If the relative phase $\phi$ between these ports is scanned, 
the count rate in either output port of the Mach-Zhender interferometer will oscillate as
$P_\mathrm{MZ}=\frac{1}{2}+\frac{1}{2}V\cos\phi$,
where $V$ is the visibility of $P_\mathrm{MZ}$:
\begin{equation}
V=\sqrt{1-v^2\Gamma^2}.
\end{equation}
Our goal is to measure this visibility $V$ experimentally when different phases are present in the Sagnac interferometer, and use this information to draw conclusions about the commutativity of the phases in our experiment via $\Gamma$. 
As we will see, if we perform two different measurements, each with different phases in the interferometer (two different values of $\Gamma$), we can alltogher avoid needing to know $v$ the visibility of the Sagnac interferometer.

The choice of test phases is important for discovering potential quanternionic phases.
In his proposal, Peres suggested a neutron interferometry experiment that used materials with complex scattering amplitudes, arguing that such materials would be more likely to have a quaternionic component.
Based on this, Kaiser \textit{et al} used one phase shift with a positive  scattering amplitude and one with a negative scattering amplitude \cite{Kaiser1984}.
Similarly, we choose two optical materials with very different phase responses:
one material with a positive refractive index, and one with a negative refractive index.  We use a standard liquid-crystal phase retarder to provide a uniform, low-optical-loss phase shift as our first positive phase.

For our second phase we use an artificial nanostructured metamaterial.
These materials have recently been used to probe several exciting quantum phenomena \cite{asano2015distillation,altewischer2002plasmon}.
We designed our metamaterial to have a negative refractive index, and thus apply a negative phase.
Achieving this requires both the real part of permittivity and permeability to be negative.
We obtain this at optical frequencies with a fishnet optical metamaterial which integrates two types of structures together -- one with a negative permittivity, and one with a negative permeability.
See the Appendix for more information.

\section*{Results}

\begin{figure*}
\begin{center} 
\includegraphics[width = \textwidth]{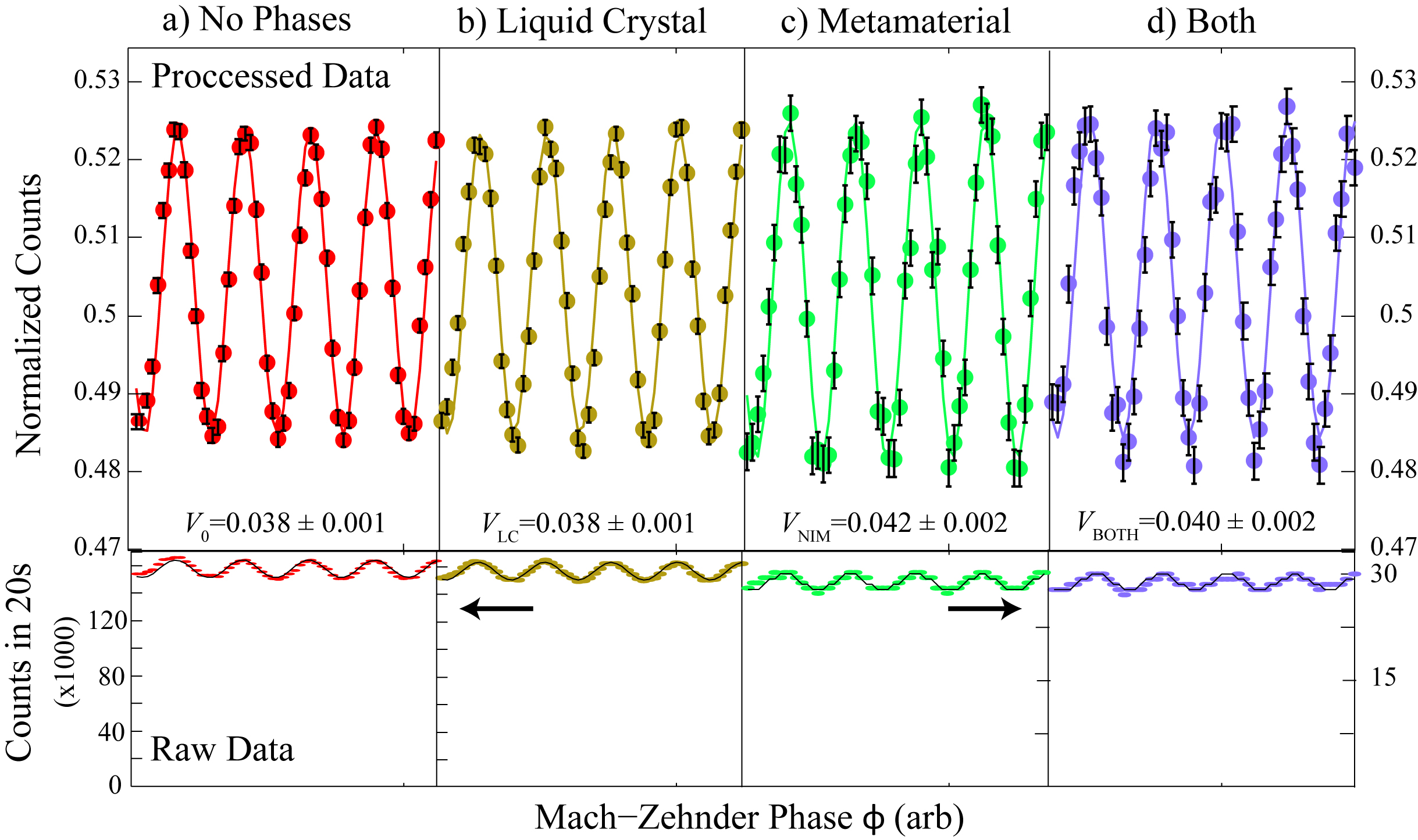}
\label{rawData}
\end{center}
\caption{\textbf{Representative Interferograms of the Mach-Zehnder Interferometer}.  All of these data are counts plotted versus the phase of the Mach-Zehnder interferometer.
The data in the upper row are the count rate of photons exiting port one (at D1) normalized to the sum of the counts out our both ports.
The lower row shows the same data without normalization.
Measurements for four cases are shown: (\textbf{a}) no phases inside of Sagnac loop, (\textbf{b}) only a positive phase (using the liquid crystal), (\textbf{c}) only a negative phase (using the negative-index metamaterial), and (\textbf{d}) both phases engaged. The visibilities of the data shown in panels c) and d) are equal within error. 
The unnormalized data (lower row) presented a) and b) uses the scale bar on the left, while the unnormalized data of panels c) and d) uses the scale bar on the right.  
The decreased count rate is due to the 13\% transmission of the metamaterial.}
\end{figure*}

\begin{figure*}
\begin{center} 
\includegraphics[width = \textwidth]{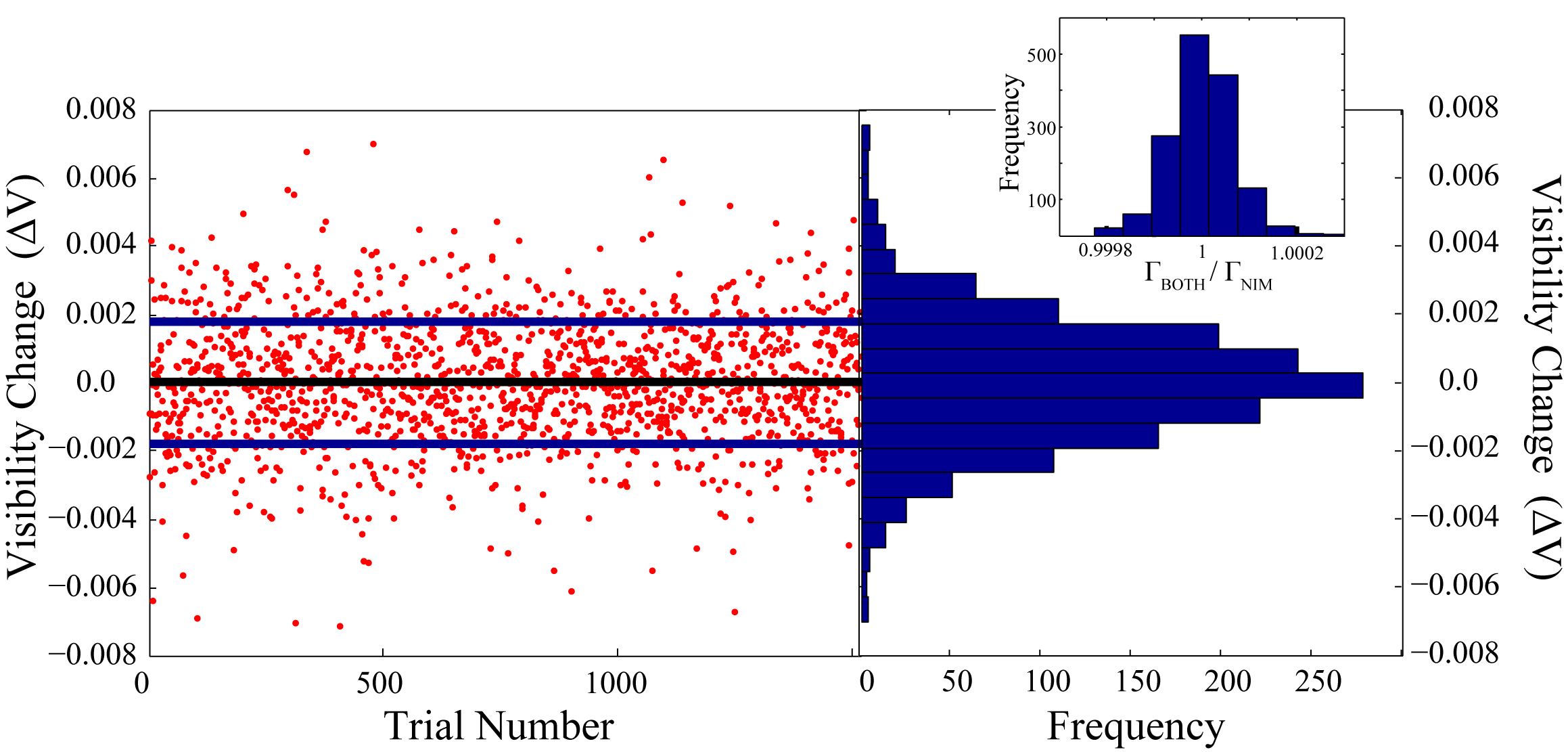}
\label{repeatedData}
\end{center}
\caption{\textbf{Results for repeated runs of the experiment.} (\textbf{a}) Each point corresponds to one run of the experiment, consisting of turning the liquid crystal off and measuring the visibility of the Mach-Zhender interferometer, followed by turning the liquid crystal on and remeasuring the visibility.
The difference between these two visibilities for each run is plotted here for data from each of the two ports of the Mach-Zehnder interferometer.
A total of 761 experimental runs were made, resulting in 1522 values of  $\Delta V$.
The black line marks the mean of all of the points, and the blue lines the standard deviation. (\textbf{b}) A histogram of data plotted in panel a). 
\textbf{Inset:} A histogram of the values of $\frac{\Gamma_\mathrm{BOTH}}{\Gamma_\mathrm{NIM}}$, computed from the data in panel (a). 
The mean value of this distribution is used to compute a phase difference between photons seeing the liquid-crystal phase retarder before of after the metamaterial.
A mean value of this distribution that is not equal to $1$ would indicate some form of non-commutativity.}
\end{figure*}

A sketch of our experimental implementation is presented in Fig. \ref{expt}.  
We send heralded single photons (see the Appendix) into a Sagnac interferometer.
The Sagnac interferometer has two output modes, labelled B and D in Fig. \ref{expt}.
CQM predicts that B is the bright port, and D is the dark port.
After exiting the Sagnac interferometer, photons in mode B reflect off of BS1, and those in mode D reflect off of BS2.
Beamsplitter BS2 is used to reflect mode D so that both modes experience the same attenuation, as this yields the highest visibility interference.
The two modes then interfere at BS4.
To ensure high-visibility interference, the input light is polarized with polarizer P1, and two final polarizers P2 and P3 (aligned to P1) are placed before the fiber couplers.
Finally, both modes are coupled into single-mode fiber for spatial filtering.

To measure the interference between the $B$ and $D$ modes, BS2 is mounted on a piezo-actuated translation stage to scan the phase $\phi$.
Because the visibility of the Sagnac interferometer is not perfect ($v<1$), we observe interference even without any phases in the Sagnac.
This reference signal is shown in the bottom panel of Fig. \ref{rawData}a.
In this graph, the counts registered at detector D1 are plotted versus the position of the translation stage.
We also collect the photons exiting the other port, at detector D2 to normalize the data;
these normalized data are plotted in the upper panel of Fig. \ref{rawData}a.
We extract the visibility of the normalized curve by fitting the data, as described in the Appendix, and we find that the visibility is $V_o=0.038\pm 0.001$.
The error bars are determined from the uncertainty of the fit parameters.

After characterizing our setup with no additional internal phases in the Sagnac interferometer we must characterize the individual effect of each of the two phases. 
We first turn on liquid-crystal phase retarder (LC) by applying a voltage that results in an effective phase of $\pi$ rad (see Fig. \ref{cartoon}d for the details of our LC).
A resulting interference signal is shown in Fig. \ref{rawData}b.
On a single run, turning the LC on does not introduce any measurable effects: the visibility is still $V_\mathrm{LC}=0.038\pm 0.001$. 
To reduce the influence of statistical fluctuations, this measurement is repeated 402 times.
This minimizes the effects of long term noise, since each run is faster than any observable fluctuation. 
We find that the LC produces an average visibility difference of  $\Delta_\mathrm{LC}=V_\mathrm{LC}-V_o= 0.002\pm 0.003$.
{This result is consistent with $0$, so we see turning on the LC has essentially no effect on our experimental apparatus.
This confirms that the visibility of the Sagnac interferometer $v$ is independent of the LC, and we can use this result to bound the systematic error induced by the LC.}
Note that these two measurements (presented in Fig. \ref{rawData}a and \ref{rawData}b) are only used to characterize our apparatus.
 
Next, we study the second phase: a negative phase shift of $-\pi$, which is induced by inserting the negative index metamaterial (NIM) into the Sagnac interferometer.
The results of the negative-phase characterization are presented in Fig. \ref{cartoon}c.
Data with the NIM inserted and the LC phase set to 0 rad are shown in Fig. \ref{rawData}c. 
The NIM has a transmission of $13\%$ at 790 nm, which is evident in the lower count rate of the raw data.
We find that inserting the NIM marginally decreases the visibility of the Sagnac interferometer, leading to an  increased Mach-Zehnder visibility of $V_\mathrm{NIM}=0.042\pm0.002$.
This visibility increase occurs because inserting the NIM slightly degrades or shifts the spatial modes inside the Sagnac interferometer.
We believe that this increase in visibility is rather a systematic error, and not the quaternionic effect that we are interested in.

To observe an effect due to potential non-commutativity we only need study how visbilities change in response to different phases in the interferometer.
Since the systematic error of the LC phase is much smaller than the error caused by inserting the NIM, we leave the NIM inserted and compare the visibility when the LC phase is set to $0$ rad and $\pi$ rad.
This allows us to neglect the larger systematic error of inserting the NIM.

Data with both the NIM inserted and LC phase set to $\pi$ are shown in Fig. \ref{rawData}d, and have a visibility of $V_\mathrm{BOTH}=0.040\pm 0.002$.
We need to compare this to the data presented in Fig. \ref{rawData}c.
On a single run the two visibilities are equal within experimental error, i.e. $V_\mathrm{NIM}=V_\mathrm{BOTH}$.
This already indicates that the two phases commute.

To decrease our statistical errors to the level of the LC systematic error, we repeat this experiment.
We first set the LC to first to $0$ rad and then $\pi$ rad a total of $761$ times, while leaving the NIM inserted the entire time.
In other words, we generate the data presented in Fig. \ref{rawData}c and \ref{rawData}d many times.
For each run we measure $\Delta V=V_\mathrm{BOTH}-V_\mathrm{NIM}$ for the data collected out of both ports at detecctors D1 and D2, yielding a total of $1522$ values for $\Delta V$.
These data are shown in Fig. \ref{repeatedData}a, and
a histogram of these results in presented in Fig. \ref{repeatedData}b.
Notice that on a given trial, $V_\mathrm{NIM}$ can appear larger than $V_\mathrm{BOTH}$, leading to a negative value.
However, within error $V_\mathrm{NIM}$ and $V_\mathrm{BOTH}$ are equal for most trials. 
To be more precise, we examine the mean value of this distribution $\Delta V=0.0006\pm 0.005$.
This is consistent with zero, and it indicates that the two phases in our experiment commute with a very high precision.
The statistical error on $\Delta V$ is $0.005$, which is slightly larger than the systematic error coming from turning on the LC.

\section*{Discussion}

As a final step we convert our visibility change into a figure of merit which provides more insight into the physical meaning of our results, and allows us to compare our result to a previous neutron interferometry experiment\cite{Kaiser1984}.
Although we should point out that the deviation from CQM could be different for neutrons and photons.
In the neutron experiment it was found that two interference patterns (each created with two phases shifters inserted in either order) were shifted by less than $0.3^\circ$.
Then, since each phase shifter imparted a phase on the order of $10,000^\circ$, they concluded that any quaternionic contribution must be less than 1 part in $30,000$.
However, this assumes that the quaterionic phase is linearly proportional to total phase---there is no such requirement in QQM  (see the Appendix). 
In fact, the quaternionic phase could be completely independent of the standard quantum phase.
Thus only the absolute deviation from CQM's predictions is relevant to the quaternionic non-commutativity, and relevant bound from the previous work is $0.3^\circ$.

To start this conversion, we extract the ratio of $\Gamma$ when both phases are activated to $\Gamma$ when only the NIM is inside the Sagnac, $\Gamma_\mathrm{BOTH}/\Gamma_\mathrm{NIM}$, from the following definition
\begin{equation}\label{eq:GAMMAmain}
\frac{\Gamma_\mathrm{BOTH}}{\Gamma_\mathrm{NIM}} = \sqrt{\frac{1-V_\mathrm{BOTH}^2}{1-V_\mathrm{NIM}^2}}.
\end{equation}
Here, $\Gamma_\mathrm{BOTH}$ is defined in Eq. \ref{eq:gammaBOTHfull}, and  $\Gamma_\mathrm{NIM}$ is defined in Eq. \ref{eq:gammaNIM} of the Appendix.
If this ratio deviates from 1, then there must be some non-commutativity.
We can further convert this ratio into a phase shift simply as $\theta=\mathrm{acos}(\frac{\Gamma_\mathrm{BOTH}}{\Gamma_\mathrm{NIM}})$.
See the Appendix for more details.
We use Eq. \ref{eq:GAMMAmain} to compute $\Gamma_\mathrm{BOTH}/\Gamma_\mathrm{NIM}$ for every data point, the resulting distribution is shown in the inset of Fig. \ref{repeatedData}b.
From the mean of this distribution we find $\Gamma_\mathrm{BOTH}/\Gamma_\mathrm{NIM}=1$ with a precision of $2\times 10^{-7}$, i.e. $\Gamma_\mathrm{BOTH}/\Gamma_\mathrm{NIM}=0.99999999\pm 2\times 10^{-7}$.
Converting this a phase shift yields a bound of $\theta=0.03^\circ$---one order of magnitude smaller than the previous experiment.

In light of this analysis, our result can be seen as an extremely high-precision measurement of a phase shift between the two modes of the Sagnac interferometer. 
In principle, such a phase shift could arise from other effects, even in a common-path Sagnac interferometer such as ours.
However, in our estimation, all of these potential phase shifts are orders of magnitude smaller than our null result.
For example, given the geometry of our interferometer, the rotation of the Earth could lead to a phase shift of at most $10^{-4}$ degrees;
Faraday effects caused by the Earth's magnetic field would be even smaller.
Since they would be constant, all such phase shifts would present themselves as a reduced visibility of the Sagnac interferometer.
Although we find an imperfect visibility of the Sagnac interferometer, we attribute this to a slight mismatch between the spatial modes of the Sagnac interferometer.
Given the polarizers before and after the interferometer, polarization mismatch between the two modes, although possible, is very small.
We observed that this effect is smaller than the the spatial mismatch of the two modes.
Moreover, these effects lead to an systematic decrease in the visibility of the Sagnac interferometer, and our data analysis accounts for this.

Our work is the first direct search for the presence of quaternions in optics \cite{klein1988}.
{It is enabled by the combination of a novel negative-index metamaterial with standard optical photonic technology.}
We have tightened the previous bound on quaternionic non-commutativty, finding that a quaternionic description of quantum mechanics is not required for our experiment.
Our bound is relevant to any work generalizing quantum mechanics.
It is still very important to continue the search for effects predicted by new, generalized quantum theories.
Further tests of QQM could be performed in optics using other methods to apply phases, or in other physical systems using molecular, electron, or other matter-wave interferometers.

\noindent\textbf{Acknowledgements}\\
We thank \v{C}aslav Brukner and Gregor Weihs for stimulating discussions, and T.
R\"{o}gelsperger for assisting with the figures.\\
We acknowledge support from the European Commission, QUILMI (No. 295293), EQUAM (No. 323714), PICQUE (No. 608062), GRASP (No.613024), QUCHIP (No.641039), and RAQUEL (No. 323970); 
the Austrian Science Fund (FWF) through START (Y585-N20), the doctoral programme CoQuS, and Individual Project (No. 2462);
the Vienna Science and Technology Fund (WWTF, grant ICT12-041);
the United States Air Force Office of Scientific Research (FA8655-11-1-3004); 
and the Foundational Questions Institute (FQXi). L.M.P. acknowledges partial support from CONACYT-Mexico.
L.A.R was partially funded by a Natural Sciences and Engineering Research Council of Canada (NSERC) Postdoctoral Fellowship.
Z. J. W., K. O. and X. Z. are supported by the U.S. Department of Energy, Office of Science, Basic Energy Sciences, Materials Sciences and Engineering Division under contract no. DE-AC02-05CH11231.

\bibliography{quaternion_paper10_arXiv} 

\begin{thebibliography}{10}
\expandafter\ifx\csname url\endcsname\relax
  \def\url#1{\texttt{#1}}\fi
\expandafter\ifx\csname urlprefix\endcsname\relax\def\urlprefix{URL }\fi
\providecommand{\bibinfo}[2]{#2}
\providecommand{\eprint}[2][]{\url{#2}}

\bibitem{Einstein1935}
\bibinfo{author}{Einstein, A.}, \bibinfo{author}{Podolsky, B.} \&
  \bibinfo{author}{Rosen, N.}
\newblock \bibinfo{title}{Can quantum-mechanical description of physical
  reality be considered complete?}
\newblock \emph{\bibinfo{journal}{Phys. Rev.}} \textbf{\bibinfo{volume}{47}},
  \bibinfo{pages}{777--780} (\bibinfo{year}{1935}).

\bibitem{bell1964}
\bibinfo{author}{Bell, J.~S.}
\newblock \bibinfo{title}{On the {E}instein-{P}odolsky-{R}osen paradox}.
\newblock \emph{\bibinfo{journal}{Physics}} \textbf{\bibinfo{volume}{1}},
  \bibinfo{pages}{195--200} (\bibinfo{year}{1964}).

\bibitem{weihs1998}
\bibinfo{author}{Weihs, G.}, \bibinfo{author}{Jennewein, T.},
  \bibinfo{author}{Simon, C.}, \bibinfo{author}{Weinfurter, H.} \&
  \bibinfo{author}{Zeilinger, A.}
\newblock \bibinfo{title}{Violation of {B}ell's inequality under strict
  einstein locality conditions}.
\newblock \emph{\bibinfo{journal}{Phys. Rev. Lett.}}
  \textbf{\bibinfo{volume}{81}}, \bibinfo{pages}{5039--5043}
  (\bibinfo{year}{1998}).

\bibitem{Christensen2013}
\bibinfo{author}{Christensen, B.~G.} \emph{et~al.}
\newblock \bibinfo{title}{Detection-loophole-free test of quantum nonlocality,
  and applications}.
\newblock \emph{\bibinfo{journal}{Phys. Rev. Lett.}}
  \textbf{\bibinfo{volume}{111}}, \bibinfo{pages}{130406}
  (\bibinfo{year}{2013}).

\bibitem{giustina2013}
\bibinfo{author}{Giustina, M.} \emph{et~al.}
\newblock \bibinfo{title}{Bell violation using entangled photons without the
  fair-sampling assumption}.
\newblock \emph{\bibinfo{journal}{Nature}} \textbf{\bibinfo{volume}{497}},
  \bibinfo{pages}{227--230} (\bibinfo{year}{2013}).

\bibitem{hensen2015loophole}
\bibinfo{author}{Hensen, B.} \emph{et~al.}
\newblock \bibinfo{title}{Loophole-free {B}ell inequality violation using
  electron spins separated by 1.3 kilometres}.
\newblock \emph{\bibinfo{journal}{Nature}} \textbf{\bibinfo{volume}{526}},
  \bibinfo{pages}{682--686} (\bibinfo{year}{2015}).

\bibitem{giustina2015significant}
\bibinfo{author}{Giustina, M.} \emph{et~al.}
\newblock \bibinfo{title}{Significant-loophole-free test of {B}ell's theorem
  with entangled photons}.
\newblock \emph{\bibinfo{journal}{Physical review letters}}
  \textbf{\bibinfo{volume}{115}}, \bibinfo{pages}{250401}
  (\bibinfo{year}{2015}).

\bibitem{shalm2015strong}
\bibinfo{author}{Shalm, L.~K.} \emph{et~al.}
\newblock \bibinfo{title}{Strong loophole-free test of local realism}.
\newblock \emph{\bibinfo{journal}{Physical review letters}}
  \textbf{\bibinfo{volume}{115}}, \bibinfo{pages}{250402}
  (\bibinfo{year}{2015}).

\bibitem{Bialynicki1976}
\bibinfo{author}{Bialynicki-Birula, I.} \& \bibinfo{author}{Mycielski, J.}
\newblock \bibinfo{title}{{Nonlinear wave mechanics}}.
\newblock \emph{\bibinfo{journal}{Annals of Physics}}
  \textbf{\bibinfo{volume}{100}}, \bibinfo{pages}{62--93}
  (\bibinfo{year}{1984}).

\bibitem{Weinberg1989}
\bibinfo{author}{Weinberg, S.}
\newblock \bibinfo{title}{{Testing Quanutm Mechanics}}.
\newblock \emph{\bibinfo{journal}{Annals of Physics}}
  \textbf{\bibinfo{volume}{194}}, \bibinfo{pages}{336--386}
  (\bibinfo{year}{1989}).

\bibitem{grw}
\bibinfo{author}{Ghirardi, G.}, \bibinfo{author}{Rimini, A.} \&
  \bibinfo{author}{Weber, T.}
\newblock \bibinfo{title}{Unified dynamics for microscopic and macroscopic
  systems}.
\newblock \emph{\bibinfo{journal}{Physical Review D}}
  \textbf{\bibinfo{volume}{34}}, \bibinfo{pages}{470} (\bibinfo{year}{1986}).

\bibitem{Sorkin1994}
\bibinfo{author}{Sorkin, R.~D.}
\newblock \bibinfo{title}{Quantum mechanics as quantum measure theory}.
\newblock \emph{\bibinfo{journal}{Modern Physics Letters A}}
  \textbf{\bibinfo{volume}{09}}, \bibinfo{pages}{3119--3127}
  (\bibinfo{year}{1994}).

\bibitem{dakic2014density}
\bibinfo{author}{Daki{\'c}, B.}, \bibinfo{author}{Paterek, T.} \&
  \bibinfo{author}{Brukner, {\v{C}}.}
\newblock \bibinfo{title}{Density cubes and higher-order interference
  theories}.
\newblock \emph{\bibinfo{journal}{New Journal of Physics}}
  \textbf{\bibinfo{volume}{16}}, \bibinfo{pages}{023028}
  (\bibinfo{year}{2014}).

\bibitem{finkelstein1962}
\bibinfo{author}{Finkelstein, D.}, \bibinfo{author}{Jauch, J.~M.},
  \bibinfo{author}{Schiminovich, S.} \& \bibinfo{author}{Speiser, D.}
\newblock \bibinfo{title}{Foundations of quaternion quantum mechanics}.
\newblock \emph{\bibinfo{journal}{Journal of mathematical physics}}
  \textbf{\bibinfo{volume}{3}}, \bibinfo{pages}{207--220}
  (\bibinfo{year}{1962}).

\bibitem{adler1995quaternionic}
\bibinfo{author}{Adler, S.~L.}
\newblock \emph{\bibinfo{title}{Quaternionic quantum mechanics and quantum
  fields}} (\bibinfo{publisher}{Oxford Univ. Press}, \bibinfo{year}{1995}).

\bibitem{Horwitz1996}
\bibinfo{author}{Horwitz, L.}
\newblock \bibinfo{title}{Hypercomplex quantum mechanics}.
\newblock \emph{\bibinfo{journal}{Foundations of Physics}}
  \textbf{\bibinfo{volume}{26}}, \bibinfo{pages}{851--862}
  (\bibinfo{year}{1996}).

\bibitem{Kanatchikov1999}
\bibinfo{author}{Kanatchikov, I.~V.}
\newblock \bibinfo{title}{{D}e {D}onder-{W}eyl theory and a hypercomplex
  extension of quantum mechanics to field theory}.
\newblock \emph{\bibinfo{journal}{Reports on Mathematical Physics}}
  \textbf{\bibinfo{volume}{43}}, \bibinfo{pages}{157 -- 170}
  (\bibinfo{year}{1999}).

\bibitem{baez2012division}
\bibinfo{author}{Baez, J.~C.}
\newblock \bibinfo{title}{Division algebras and quantum theory}.
\newblock \emph{\bibinfo{journal}{Foundations of Physics}}
  \textbf{\bibinfo{volume}{42}}, \bibinfo{pages}{819--855}
  (\bibinfo{year}{2012}).

\bibitem{Kaiser1984}
\bibinfo{author}{Kaiser, H.}, \bibinfo{author}{George, E.} \&
  \bibinfo{author}{Werner, S.}
\newblock \bibinfo{title}{{Neutron interferometric search for quaternions in
  quantum mechanics}}.
\newblock \emph{\bibinfo{journal}{Physical Review A}}
  \textbf{\bibinfo{volume}{29}}, \bibinfo{pages}{2276--2279}
  (\bibinfo{year}{1984}).

\bibitem{klein1988}
\bibinfo{author}{Klein, A.}
\newblock \bibinfo{title}{Schr{\"o}dinger inviolate: neutron optical searches
  for violations of quantum mechanics}.
\newblock \emph{\bibinfo{journal}{Physica B+ C}}
  \textbf{\bibinfo{volume}{151}}, \bibinfo{pages}{44--49}
  (\bibinfo{year}{1988}).

\bibitem{Sinha2010}
\bibinfo{author}{Sinha, U.}, \bibinfo{author}{Couteau, C.},
  \bibinfo{author}{Jennewein, T.}, \bibinfo{author}{Laflamme, R.} \&
  \bibinfo{author}{Weihs, G.}
\newblock \bibinfo{title}{{Ruling out multi-order interference in quantum
  mechanics.}}
\newblock \emph{\bibinfo{journal}{Science (New York, N.Y.)}}
  \textbf{\bibinfo{volume}{329}}, \bibinfo{pages}{418--21}
  (\bibinfo{year}{2010}).

\bibitem{garner2014quantum}
\bibinfo{author}{Garner, A.~J.}, \bibinfo{author}{M{\"u}ller, M.~P.} \&
  \bibinfo{author}{Dahlsten, O.~C.}
\newblock \bibinfo{title}{The quantum bit from relativity of simultaneity on an
  interferometer}.
\newblock \emph{\bibinfo{journal}{arXiv preprint arXiv:1412.7112}}
  (\bibinfo{year}{2014}).

\bibitem{Peres1979}
\bibinfo{author}{Peres, A.}
\newblock \bibinfo{title}{{Proposed Test for Complex versus Quaternion Quantum
  Theory}}.
\newblock \emph{\bibinfo{journal}{Physical Review Letters}}
  \textbf{\bibinfo{volume}{42}}, \bibinfo{pages}{683--686}
  (\bibinfo{year}{1979}).

\bibitem{dolling2007negative}
\bibinfo{author}{Dolling, G.}, \bibinfo{author}{Wegener, M.},
  \bibinfo{author}{Soukoulis, C.~M.} \& \bibinfo{author}{Linden, S.}
\newblock \bibinfo{title}{Negative-index metamaterial at 780 nm wavelength}.
\newblock \emph{\bibinfo{journal}{Optics letters}}
  \textbf{\bibinfo{volume}{32}}, \bibinfo{pages}{53--55}
  (\bibinfo{year}{2007}).

\bibitem{shalaev2007}
\bibinfo{author}{Shalaev, V.~M.}
\newblock \bibinfo{title}{Optical negative-index metamaterials}.
\newblock \emph{\bibinfo{journal}{Nature photonics}}
  \textbf{\bibinfo{volume}{1}}, \bibinfo{pages}{41--48} (\bibinfo{year}{2007}).

\bibitem{yao2008}
\bibinfo{author}{Yao, J.} \emph{et~al.}
\newblock \bibinfo{title}{Optical negative refraction in bulk metamaterials of
  nanowires}.
\newblock \emph{\bibinfo{journal}{Science}} \textbf{\bibinfo{volume}{321}},
  \bibinfo{pages}{930--930} (\bibinfo{year}{2008}).

\bibitem{Zhang2011}
\bibinfo{author}{Liu, Y.} \& \bibinfo{author}{Zhang, X.}
\newblock \bibinfo{title}{Metamaterials: a new frontier of science and
  technology}.
\newblock \emph{\bibinfo{journal}{Chem. Soc. Rev.}}
  \textbf{\bibinfo{volume}{40}}, \bibinfo{pages}{2494--2507}
  (\bibinfo{year}{2011}).

\bibitem{Birkhoff1936}
\bibinfo{author}{Birkhoff, G.} \& \bibinfo{author}{Neumann, J.~V.}
\newblock \bibinfo{title}{The logic of quantum mechanics}.
\newblock \emph{\bibinfo{journal}{Annals of Mathematics}}
  \textbf{\bibinfo{volume}{37}}, \bibinfo{pages}{pp. 823--843}
  (\bibinfo{year}{1936}).

\bibitem{hamilton1844}
\bibinfo{author}{Hamilton, W.~R.}
\newblock \bibinfo{title}{On quaternions; or on a new system of imaginaries in
  algebra}.
\newblock \emph{\bibinfo{journal}{The London, Edinburgh, and Dublin
  Philosophical Magazine and Journal of Science}}
  \textbf{\bibinfo{volume}{25}}, \bibinfo{pages}{10--13}
  (\bibinfo{year}{1844}).

\bibitem{brumby1996}
\bibinfo{author}{Brumby, S.~P.} \& \bibinfo{author}{Joshi, G.~C.}
\newblock \bibinfo{title}{Experimental status of quaternionic quantum
  mechanics}.
\newblock \emph{\bibinfo{journal}{Chaos, Solitons \& Fractals}}
  \textbf{\bibinfo{volume}{7}}, \bibinfo{pages}{747--752}
  (\bibinfo{year}{1996}).

\bibitem{popescu1994quantum}
\bibinfo{author}{Popescu, S.} \& \bibinfo{author}{Rohrlich, D.}
\newblock \bibinfo{title}{Quantum nonlocality as an axiom}.
\newblock \emph{\bibinfo{journal}{Foundations of Physics}}
  \textbf{\bibinfo{volume}{24}}, \bibinfo{pages}{379--385}
  (\bibinfo{year}{1994}).

\bibitem{paterek2010theories}
\bibinfo{author}{Paterek, T.}, \bibinfo{author}{Daki{\'c}, B.} \&
  \bibinfo{author}{Brukner, {\v{C}}.}
\newblock \bibinfo{title}{Theories of systems with limited information
  content}.
\newblock \emph{\bibinfo{journal}{New Journal of Physics}}
  \textbf{\bibinfo{volume}{12}}, \bibinfo{pages}{053037}
  (\bibinfo{year}{2010}).

\bibitem{barrett2007information}
\bibinfo{author}{Barrett, J.}
\newblock \bibinfo{title}{Information processing in generalized probabilistic
  theories}.
\newblock \emph{\bibinfo{journal}{Physical Review A}}
  \textbf{\bibinfo{volume}{75}}, \bibinfo{pages}{032304}
  (\bibinfo{year}{2007}).

\bibitem{kocsis2011}
\bibinfo{author}{Kocsis, S.} \emph{et~al.}
\newblock \bibinfo{title}{Observing the average trajectories of single photons
  in a two-slit interferometer}.
\newblock \emph{\bibinfo{journal}{Science}} \textbf{\bibinfo{volume}{332}},
  \bibinfo{pages}{1170--1173} (\bibinfo{year}{2011}).

\bibitem{rozema2012}
\bibinfo{author}{Rozema, L.~A.} \emph{et~al.}
\newblock \bibinfo{title}{Violation of {H}eisenberg's measurement-disturbance
  relationship by weak measurements}.
\newblock \emph{\bibinfo{journal}{Physical Review Letters}}
  \textbf{\bibinfo{volume}{109}}, \bibinfo{pages}{100404}
  (\bibinfo{year}{2012}).

\bibitem{Shadbolt2014}
\bibinfo{author}{Shadbolt, P.}, \bibinfo{author}{Mathews, J. C.~F.},
  \bibinfo{author}{Laing, A.} \& \bibinfo{author}{O'Brien, J.~L.}
\newblock \bibinfo{title}{{Testing foundations of quantum mechanics with
  photons}}.
\newblock \emph{\bibinfo{journal}{Nature Physics}}
  \textbf{\bibinfo{volume}{10}}, \bibinfo{pages}{278--286}
  (\bibinfo{year}{2014}).

\bibitem{asano2015distillation}
\bibinfo{author}{Asano, M.} \emph{et~al.}
\newblock \bibinfo{title}{Distillation of photon entanglement using a plasmonic
  metamaterial}.
\newblock \emph{\bibinfo{journal}{arXiv preprint arXiv:1507.07948}}
  (\bibinfo{year}{2015}).

\bibitem{altewischer2002plasmon}
\bibinfo{author}{Altewischer, E.}, \bibinfo{author}{Van~Exter, M.} \&
  \bibinfo{author}{Woerdman, J.}
\newblock \bibinfo{title}{Plasmon-assisted transmission of entangled photons}.
\newblock \emph{\bibinfo{journal}{Nature}} \textbf{\bibinfo{volume}{418}},
  \bibinfo{pages}{304--306} (\bibinfo{year}{2002}).

\bibitem{minovich2010tilted}
\bibinfo{author}{Minovich, A.} \emph{et~al.}
\newblock \bibinfo{title}{Tilted response of fishnet metamaterials at
  near-infrared optical wavelengths}.
\newblock \emph{\bibinfo{journal}{Physical Review B}}
  \textbf{\bibinfo{volume}{81}}, \bibinfo{pages}{115109}
  (\bibinfo{year}{2010}).

\end{thebibliography}

\cleardoublepage
\section*{Appendix}

\subsection{Single-Photon Source}
Our single-photon source is based on a Sagnac interferometer, commonly used to create polarization-entangled photon pairs, but we generate photon pairs in a separable polarization state.
Our Sagnac loop is built using a dual-wavelength polarizing beamsplitter (dPBS) and two mirrors. 
A type-II collinear periodically-poled Potassium Titanyl
Phosphate (PPKTP) crystal of length 20 mm is placed inside the loop and pumped by a 23.7 mW diode laser centered at 395 nm. 
This results in photon pairs at a degenerate wavelength of 790 nm. 
The pump beam polarization is set to horizontal in order to generate the down-converted photons in a separable polarization state $\ket{H}\ket{V}$. 
The dichroic mirror (DM) transmits the pump beam and reflects the down-converted photons, and the half wave plate (HWP) and quarter waveplate (QWP) are used to adjust the polarization of the pump beam. 
Long (LP) and narrow band (BP) pass filters block the pump beam and select the desired down-converted wavelength. 
Polarizers are aligned to transmit only down-converted photons with the desired polarization. 
After this, the down-converted photon pairs are coupled into single-mode fibres (SMF), and one photon from the pair is used as a herald while the other single photon is sent to the rest of the experiment using a fibre collimator (FC). 

\subsection{Theoretical Treatment of the Sagnac Interferometer}
Here we derive the probability of a photon incident on an imperfect Sagnac interferometer to exit the ``dark port'' if two phases internal to the Sagnac interferometer do not commute.

We start with a single-photon incident on a 50:50 beamsplitter.
Ideally, given a 50:50 beamsplitter and a reflection phase of $\pi/2$, the state of a photon after reflecting is:
\begin{equation}\label{eq:state1}
(i\ket{1,0}_\mathrm{CW,CCW}+\ket{0,1}_\mathrm{CW,CCW})/\sqrt{2},
\end{equation}
where CW and CCW refer to the clockwise and counter-clockwise modes in Fig. \ref{cartoon}a, respectively.
Next, applying two phases (as in Fig. \ref{cartoon}a), represented by operators $A$ and $B$, we have
\begin{equation}\label{eq:state2}
(ABi\ket{1,0}_\mathrm{CW,CCW} + BA\ket{0,1}_\mathrm{CW,CCW})/\sqrt{2}.
\end{equation}
To be completely general we will assume that the `$i$' does not commute with $A$ and $B$.

The operators $A$ and $B$ can be represented as
\begin{equation}\label{eq:AB}
A = \alpha\mathcal{I}, 
\hspace{3ex}
B = \beta\mathcal{I}, 
\end{equation}

\noindent where $\mathcal{I}$ is the identity operator. 
In complex quantum mechanics $\alpha=e^{i \phi_A}$ and $\beta=e^{i \phi_B}$, where $\phi_A$ and $\phi_B$ are real numbers.
In this case, $\alpha$ and $\beta$ are complex numbers so $A$ and $B$ commute.
However, in quaternionic quantum mechanics the phase $\phi_A$ is generalized to vector $\{\phi_A^1,\phi_A^2,\phi_A^3\}$, where $\phi_A^1$, $\phi_A^2$, and $\phi_A^3$ are real numbers.
Then $i\phi_A$ is replaced with $i\phi_A^1+ j\phi_A^2+ k\phi_A^3$, where $\{i,j,k\}$ is a basis over the imaginary part of the quaternionic space.
With these definitions $\alpha$ and $\beta$ in Eq. \ref{eq:AB} become unit quaternions
\begin{equation}
\alpha=e^{i\phi_A^1+ j\phi_A^2+ k\phi_A^3}, 
\hspace{3ex}
\beta=e^{i\phi_B^1+ j\phi_B^2+ k\phi_B^3}
\end{equation}
Now, the operators $A$ and $B$ of Eq. \ref{eq:AB} no longer commute in general. 
In fact, $\alpha$ and $\beta$ could be even more general hyper-complex numbers, consisting of more than three imaginary components.

Next, by applying the form of the operators defined in Eq. \ref{eq:AB}, we can write state in Eq. \ref{eq:state2} as
\begin{equation}\label{eq:state3}
(\alpha\beta i\ket{1,0}_\mathrm{CW,CCW} + \beta\alpha\ket{0,1}_\mathrm{CW,CCW})/\sqrt{2}.
\end{equation}
In complex quantum mechanics, $\alpha\beta=\beta\alpha$ and the two complex numbers describe a global phase, so they have no effect on experimental outcomes.
However, if $\alpha$ and $\beta$ do not commute, the output state is
\begin{equation}\label{eq:state4}
\frac{1}{2}(\alpha\beta i + i\beta\alpha)\ket{1,0}_\mathrm{B,D} + \frac{1}{2} (i\alpha\beta i + \beta\alpha)\ket{0,1}_\mathrm{B,D}.
\end{equation}
Thus the probability for an incident photon to exit the Sagnac interferometer via the dark port (the amplitude of the second term) is
\begin{eqnarray}\label{eq:almostComm}
P_\mathrm{D}^\mathrm{ideal} &=& \frac{1}{4}|i\alpha\beta - \beta\alpha i|^2.
\end{eqnarray}
This quantifies the degree of commutativity between $\alpha$, $\beta$, and $i$. 
If $\alpha$, $\beta$, and $i$ all mutually commute it is zero.
Moreover, if $i$ commutes with $\alpha$ and $\beta$ it simply becomes the commutator of $\alpha$ and $\beta$,
as shown in Eq. \ref{eq:PdarkID} of the main text. 

We will next treat the imperfect alignment of our interferometer.
We start by writing the state from Eq. \ref{eq:state3} as a density matrix
\begin{equation}
\frac{1}{2}\left(\begin{array}{cc} 1 & \alpha\beta i {\alpha^*}{\beta^*}\\
-\beta\alpha i {\beta^*}{\alpha^*} & 1 \end{array}\right),
\end{equation}
where the $^*$ denotes the conjugate of a quaternion or complex number.
Let our Sagnac interferometer have a visibility of $v=(P_\mathrm{B}-P_\mathrm{D})/(P_\mathrm{P}+P_\mathrm{D})$, where $P_\mathrm{D}$ and $P_\mathrm{B}$ are the intensities of the dark and bright ports, respectively.
We can model this by simply scaling the coherences by $v$, as
\begin{equation}
\frac{1}{2}\left(\begin{array}{cc} 1 & v\alpha\beta i {\alpha^*}{\beta^*}\\
-v\beta\alpha i {\beta^*}{\alpha^*}  & 1 \end{array}\right).
\end{equation}
This reduced coherence can be derived by coupling the CW and CCW modes to additional modes, and then tracing out those additional modes.
This is a very general method to model imperfections since it does not require any assumptions on the types of imperfections: the CW and CCW modes could couple to additional spatial modes, temporal modes, etc.

Again, we can compute the probability to find the photon in the dark port by applying the beamsplitter transformation.
Doing so yields
\begin{equation}\label{eq:PDARK}
P_\mathrm{D}=\frac{1}{2}-\frac{v}{2}\left(1-\frac{|i\alpha\beta - \beta\alpha i|^2}{2}\right),
\end{equation}
Then the probability of the photon to exit the bright port is simply $P_\mathrm{B}=1-P_\mathrm{D}$.

\subsection{Theoretical Treatment of the Mach-Zhender Interferometer}
After the Sagnac interferometer, the bright and dark ports are interfered in our Mach-Zehnder interferometer (see Fig. \ref{cartoon}b).
Interfering two optical fields, with intensities of $P_B$ and $P_D$, on a 50:50 beamsplitter results in a signal with a visibility of.
\begin{equation}\label{eq:VMZ}
V=2\sqrt{P_\mathrm{B} P_\mathrm{D}},
\end{equation}
The same result holds if $P_B$ and $P_D$ are instead the probabilities of finding a photon in either path.
Thus, the visibility of the Mach-Zehnder interferometer with both phases inserted in the Sagnac interferometer, can be computed from $P_\mathrm{D}$ (Eq. \ref{eq:PDARK}).
After simplifying, we arrive at 
\begin{eqnarray}\label{eq:vMZ}
V_\mathrm{BOTH}=\sqrt{1-v^2\Gamma_\mathrm{BOTH}^2},
\end{eqnarray}
where
\begin{eqnarray}\label{eq:gammaBOTHfull}
\Gamma_\mathrm{BOTH}=1-\frac{1}{2}|i\alpha\beta - \beta\alpha i|^2
\end{eqnarray}
This visibility $V_\mathrm{BOTH}$ is a function of both the degree of commutativity $|i\alpha\beta - \beta\alpha i|$ and the visibility of the Sagnac interferometer $v$.
To compare to our experimental procedure imagine that we turn off the liquid-crystal phase (which we represent by $\alpha$) and leave the negative-index metamaterial inserted.
Then $\alpha$ drops out and the degree of commutativity become the commutator of $i$ and $\beta$, so Eq. \ref{eq:vMZ} becomes
\begin{eqnarray}\label{eq:vMZoff}
V_\mathrm{NIM}=\sqrt{1-v^2\Gamma_\mathrm{NIM}^2},
\end{eqnarray}
where
\begin{eqnarray}\label{eq:gammaNIM}
\Gamma_\mathrm{NIM}=1-\frac{1}{2}|[i,\beta]|^2.
\end{eqnarray}
This visibility that depends on the commutation of the negative-index metamaterial with the reflection phase inside the Sagnac interferometer, and on the visibility $v$ of the Sagnac interferometer.

By combining equations \ref{eq:vMZoff} and \ref{eq:vMZ} we arrive at a result which does not depends only on two measurable visibilities of the Mach-Zehnder interferometer, and not on the visibility $v$ of the Sagnac interferometer:
\begin{equation}\label{eq:GAMMA}
\frac{1-\frac{1}{2}|i\alpha\beta - \beta\alpha i|^2}{1-\frac{1}{2}|[i,\beta]|^2} = \sqrt{\frac{1-V_\mathrm{BOTH}^2}{1-V_\mathrm{NIM}^2}} \equiv\frac{\Gamma_\mathrm{BOTH}}{\Gamma_\mathrm{NIM}},
\end{equation}
Thus we can experimentally determine the ratio $\Gamma_\mathrm{BOTH}/\Gamma_\mathrm{NIM}$ from two visibilities of the Mach-Zehnder interferometer with and without the liquid-crystal phase turned on.
The left-hand side of Eq. \ref{eq:GAMMA} simplifies to Eq. \ref{eq:PIdark} of the main text if $i$ commutes with both $\alpha$ and $\beta$.
Notice also that if $|i\alpha\beta - \beta\alpha i|=|[i,\beta]|\neq 0$ this ratio will be one.
Thus this parameter is insensitive to a very specific type of non-commutativity where $|i\alpha\beta - \beta\alpha i|=|[i,\beta]|$.
The reason for defining the quantity $\Gamma_\mathrm{BOTH}/\Gamma_\mathrm{NIM}$ will become clear in the next section.

\subsection{Converting the Mach-Zhender Visibility Change into a Phase Change}
In this section we will derive a  figure-of-merit which both provides some physical intuition into our results, and allows us to compare our experimental precision to past work.
In the neutron interferometry experiment\cite{Kaiser1984}, two interference signals were measured (each with two phases inserted in a different order), and the phase difference between the two was used place limits on the commutativity of the phases.
In our experiment, we measure the visibility of an interference signal which is proportional to the commutator of two phases.

The signal that we monitor arises from an interference between the dark and bright output ports of the Sagnac interferometer.
As we show above, if two phases inside the Sagnac do no commute, light will leak into the dark port.
Then interfering the bright and dark modes leads to an interference signal which has a visibility given by Eq. \ref{eq:VMZ}.

Imagine that leakage into the dark port arises from of a phase shift $\theta$ between the clockwise and the counter-clockwise modes of the Sagnac.
Physically, this means that there is a different phase shift if the photon sees the metamaterial before or after the liquid-crystal.
It is straightforward to show, within CQM, that if the two modes of a Sagnac interferometer experience a phase shift $\theta$ the probabilities of the photon exiting either port become
\begin{eqnarray}\label{eq:probs}
P_\mathrm{B}=\frac{1}{2}+\frac{v}{2}\cos\theta,\\\nonumber
P_\mathrm{D}=\frac{1}{2}-\frac{v}{2}\cos\theta,
\end{eqnarray}
where $v$ is the visibility of the Sagnac interferometer.
Now substituting Eq. \ref{eq:probs} into Eq. \ref{eq:VMZ} we arrive at the visibility of the Mach-Zehnder interferometer as a function of the phase inside the Sagnac interferometer
\begin{equation}\label{eq:VTHETA}
V(\theta)=2\sqrt{1-v^2\cos^2\theta}.
\end{equation}

Experimentally, we measure two visibilities of the Mach-Zehnder interferometer, which we now attribute to a phase change in the Sagnac interferometer.
In the present picture, $V(\theta)$ and $V(0)$ are the visibilities of the Mach-Zehnder interferometer with and without a phase difference between the clockwise and counter-clockwise modes.
Thus, we will equate $V(0)$ to the visibility when only one phase is inside the Sagnac interferometer $V_\mathrm{NIM}\equiv V(0)$, and $V(\theta)$ to the visibility when both phases are in the Sagnac interferometer $V_\mathrm{BOTH}\equiv V(\theta)$.
Then we will substitute Eq. \ref{eq:VTHETA} into \ref{eq:GAMMA}, simplifying and solving for $\theta$.  
Doing this yields
\begin{equation}\label{eq:THETA}
\theta=\mathrm{acos}\left[ \sqrt{\frac{1-V_\mathrm{BOTH}^2}{1-V_\mathrm{NIM}^2}} \right]=\mathrm{acos}\left(\frac{\Gamma_\mathrm{BOTH}}{\Gamma_\mathrm{NIM}}\right).
\end{equation}
We can then understand this $\theta$ as an effective phase shift between the clockwise and counter-clockwise modes, arising from the non-commutativity of the phases.
So we see that measuring these two visibilities allows us to use Eq. \ref{eq:THETA} to convert our result into this phase.
Doing this, and using Gaussian error propagation on Eq. \ref{eq:THETA} results in $\theta<0.03^\circ$.

\subsection{Fitting to Extract Visibility}
To extract the visibility from the normalized data we fit a sinusoid to the data, and calculate the visibility from the fit parameters.
The explicit form of our fitting equation is
\begin{equation}\label{eq:fit}
A\sin^2(f x + p) + B,
\end{equation}
where $A$, $B$, $f$, and $p$ are all free parameters.
The visibility of this curve in Eq. \ref{eq:fit} in terms of the fit parameters is
\begin{equation}
\frac{A}{A+2B}.
\end{equation}
We compute the error on each visibility using Gaussian error propagation, starting with the fitting uncertainties.

\subsection{Liquid crystal retarder}
We use a commercial nematic liquid crystal cell whose molecules orient to an applied electrical field. 
We characterize the LC by placing it between two polarizing beamsplitter cubes with its optical axis at an angle of $45^\circ$. 
We then measure the light intensity transmitted through the second PBS as we vary the voltage applied to the LC. 
Since the transmitted intensity is proportional to $\frac{1}{2}(1+\cos \theta)$, where $\theta$ is the relative phase imparted by the LC, this measurement allows us to determine the relative phase (modulo $2\pi$) effected by the LC as a function of the applied voltage. 
The measured relative phase of the LC is shown in Fig \ref{cartoon}d.

\subsection{Negative-Index Metamaterial}
\begin{figure}
\begin{center} 
\includegraphics[width = .9\columnwidth]{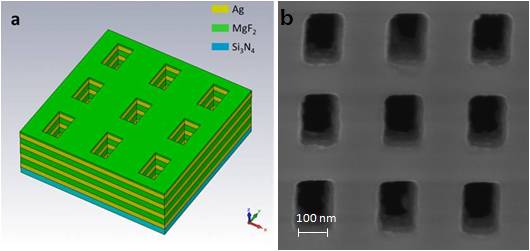}
\caption{\footnotesize \textbf{Bulk negative index metamaterial}. (\textbf{a}) Schematic of the silver (Ag)/ magnesium fluoride (MgF$_2$) multilayer fishnet metamaterials with a period of 360 nm, and a hole size of 120 nm$\times$210 nm. Negative refractive index is obtained via the coupling between the Drude-like negative permittivity background and the multiple magnetic resonances formed between each functional layer of metal/dielectric/metal nanostructures. (\textbf{b}) SEM image of the fabricated fishnet bulk NIM structure.} \label{sample}
\end{center}
\end{figure}

In our experiment, we use use a fishnet metamaterial to achieve an optical medium with a negative refractive-index.
Our meta-material consists of 7 physical layers of silver (Ag, 40 nm) and magnesium fluoride (MgF$_2$, 50 nm), with a 15 nm capping layer of MgF$_2$, see Fig. \ref{sample}. 
The nanofabrication processes included multiple steps of electron beam evaporation of Ag and MgF$_2$ materials on a 50 nm-thin low stress silicon nitride membrane, followed by Gallium-based focused ion beam lithography from the membrane side. 
A scanning electron microscope (SEM) image of the fabricated sample is shown in Fig. \ref{sample}b. 
To verify the negative phase response of the NIM, we use a spectrally and spatially resolved interferometry setup.

\begin{figure}
\begin{center} 
\includegraphics[width =1 \columnwidth]{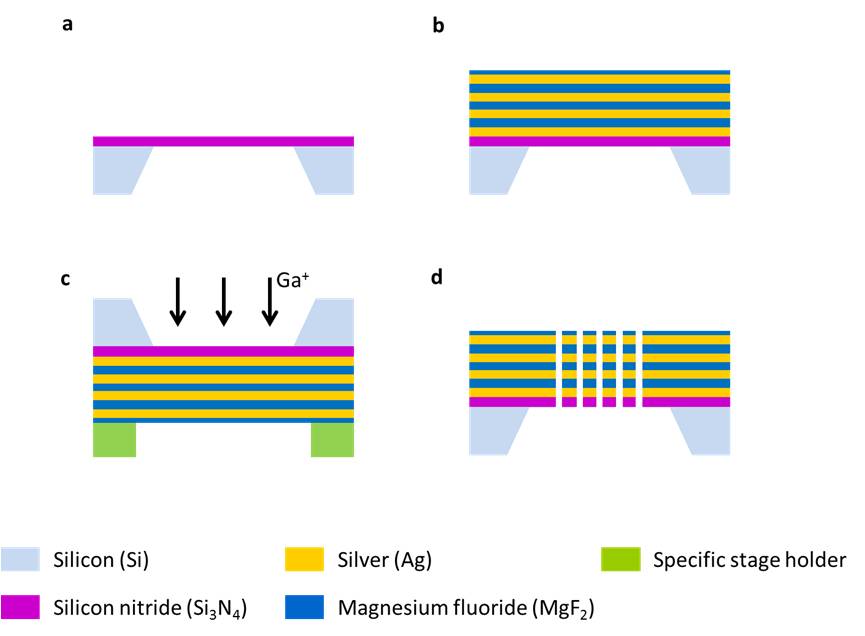}
\caption{\footnotesize \textbf{Schematic of the key steps in sample fabrication} (\textbf{a}) Fabrication of an ultra-low-stress Si$_3$N$_4$ suspended membrane. \textbf{b} Multilayer electron beam evaporation of Ag and MgF$_2$ layers without vacuum break. \textbf{(c)} Flip side mounting of sample followed by Ga$^+$ focused ion beam milling to pattern the nanostructures. (\textbf{d}) Final fishnet structure formed.
} \label{sample2}
\end{center}
\end{figure}

\begin{figure}
\begin{center} 
\includegraphics[width = .7\columnwidth]{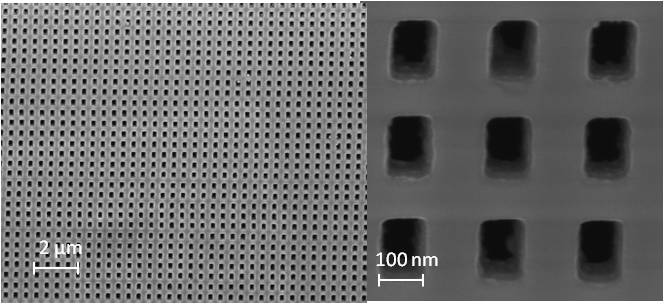}
\caption{\footnotesize \textbf{SEM images of the sample.} The top view showing the patterned nanostructures in periodic form, while the tilted view clearly displays the metal-dielectric multilayers.
} \label{sample3m}
\end{center}
\end{figure}

\noindent\textbf{Fabrication ---}
In order to attain negative phase for light passing through the sample, a suspended fishnet negative index metamaterial (NIM) is fabricated, hence avoiding any positive phase contribution from the substrate. The fabrication of the NIM starts with a suspended 50 nm ultra-low-stress silicon nitride (Si$_3$N$_4$) membrane made from standard MEMS fabrication technologies. The metal-dielectric stack is then deposited onto the Si$_3$N$_4$ membrane using layer-by-layer electron beam evaporation technique at pressure $\approx 1\times 10^{-6}$ Torr. The exact sequence of evaporation is three repetition of alternating silver (Ag, 40 nm) and magnesium fluoride (MgF$_2$, 50 nm) layers, followed by a 15 nm of MgF$_2$ as the capping layer to prevent oxidation from the top side. Next, the sample is turned upside down and mounted on a special stage holder which has a matching trench at the center. The nanostructures are milled by using focused ion-beam (FIB) from the membrane side. This is essential not just for alignment purpose, but also to reduce the optical loss caused by Ga ion penetration into the metal layers. The key fabrication steps are illustrated in Fig. \ref{sample2}. The final structure made has a slight sidewall angle along the thickness direction (Fig. \ref{sample3m}), but is previously found to have only minor influence on the negative index property. 

\noindent\textbf{Experimental characterization setup for metamaterial ---}
To measure the transmission phase change induced by the fishnet metamaterial across a broad frequency range, we built a spectrally and spatially resolved Mach-Zehnder interferometry setup. Essentially a broadband light source is split into two paths, one passing through the sample while the other serves as a reference beam, before recombining them at the input of an imaging spectrometer. The two beams interfere at different angles to produce interference fringes along the vertical axis of the imaging camera. A change in the optical path length of one of the beams will therefore cause the interference fringe to shift vertically on the image plane. By measuring the interferogram with and without the sample, and comparing them using Fourier analysis, the metamaterial induced phase change can therefore be obtained. Importantly, the phase change at different wavelengths can be captured simultaneously along the horizontal axis of the camera in a single-shot measurement.

\begin{figure}
\begin{center} 
\includegraphics[width =1\columnwidth]{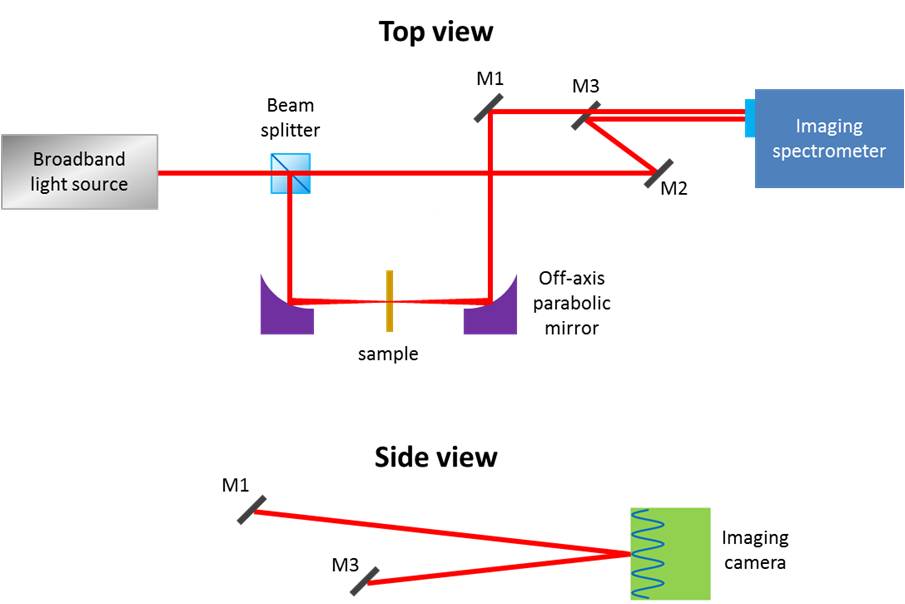}
\caption{\footnotesize \textbf{Phase measurement setup.} Interference between a beam transmitted through the sample and a reference beam allow the fringes to be formed at the imaging camera. By comparing the interferogram measured with and without the sample, the phase shift induced by the metamaterial can be determined simultaneously across a broad spectrum, limited only by the operational wavelength range of the broadband light source and the detector sensitivity.
} \label{sample4m}
\end{center}
\end{figure}

\begin{figure}
\begin{center} 
\includegraphics[width = .7\columnwidth]{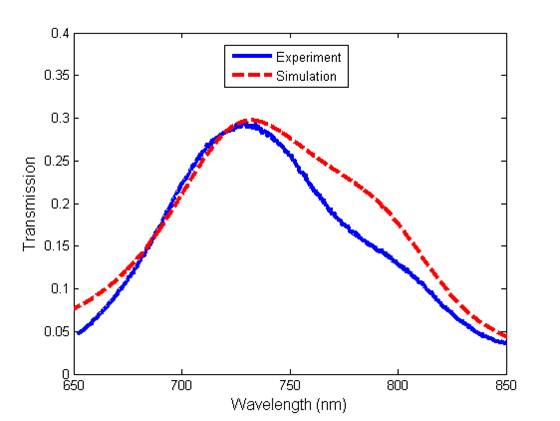}
\caption{\footnotesize \textbf{Transmission of the fishnet metamaterial} Both the measured and simulated results show a broad spectrum with relatively high transmission. Maximum transmission of 30\% is observed at ~740 nm, while 15\% transmission is measured at the desired 790 nm wavelength.} \label{sample5m}
\end{center}
\end{figure}

\noindent\textbf{Numerical design and experimental measurement of metamaterial ---}
The transmission property of the sample is important for the statistical reliability of the single photon measurement result. We performed three-dimensional full-wave finite-difference time-domain (FDTD) numerical simulations to optimize the design of the fishnet metamaterials such that the structure can attain a relatively high transmission while acquiring a negative refractive index at the desired wavelength of 790 nm. The computation is carried out using realistic material properties taking into account the dissipative loss of the silver metal used. The polarization is chosen to be $E_x$ (see Fig. \ref{sample}), which allows excitation of the anti-symmetric magnetic resonance whereby negative permeability (thus negative index) can be attained. As shown in Fig. \ref{sample5m}, several transmission peaks are observed. This multiple resonance feature arises due to the stacking of the metal/dielectric/metal layers, which lead to a low-loss broadband transmission feature at the desired wavelength range. Also shown is the experimental measurement result which matches well with the numerical design values. In particular, the experimental transmission at 790 nm is measured to be 15\%, which is among the highest reported in the visible wavelengths for bulk NIM, and is sufficient for single photon experiment.

\begin{figure}
\begin{center} 
\includegraphics[width = .7\columnwidth]{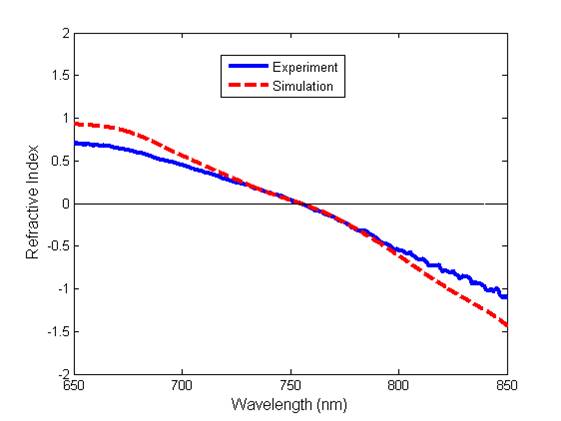}
\caption{\footnotesize \textbf{Refractive index  of the bulk fishnet metamaterial}. Both the measured and simulated refractive index values show a gradual transition from positive to negative index, with zero-crossing at around 750 nm wavelength, and an index of -0.4 at 790 nm wavelength} \label{sample6m}
\end{center}
\end{figure}

To extract the material’s effective refractive index, we simulate the transmission and reflection (both amplitude and phase) of the designed fishnet metamaterial and then reconstruct the refractive index by using the Fresnel equation. We aim for the index zero-crossing at 750 nm, above which the index will become negative. Fig. \ref{sample6m} shows the simulated refractive index from 650 nm to 850 nm wavelength range, essentially a smooth transition from positive to negative index. Experimentally, using the phase shift values measured by the setup in Fig. \ref{sample4m}, with the standard assumption of negligible multiple reflection within the structure, we could further extract its refractive index values across a broad frequency range, as depicted by the blue line in Fig.\ref{sample6m}. The trend basically agrees with the simulated results, with the slight discrepancy most likely due to the small sidewall angle of the actual fishnet structures and other fabrication-induced errors. A broadband negative index property is thus obtained from 750 nm up to 850 nm. At the wavelength of interest 790 nm, both the designed and measured refractive index is found to be -0.4. 

\begin{figure}[h]
\begin{center} 
\includegraphics[width = 1\columnwidth]{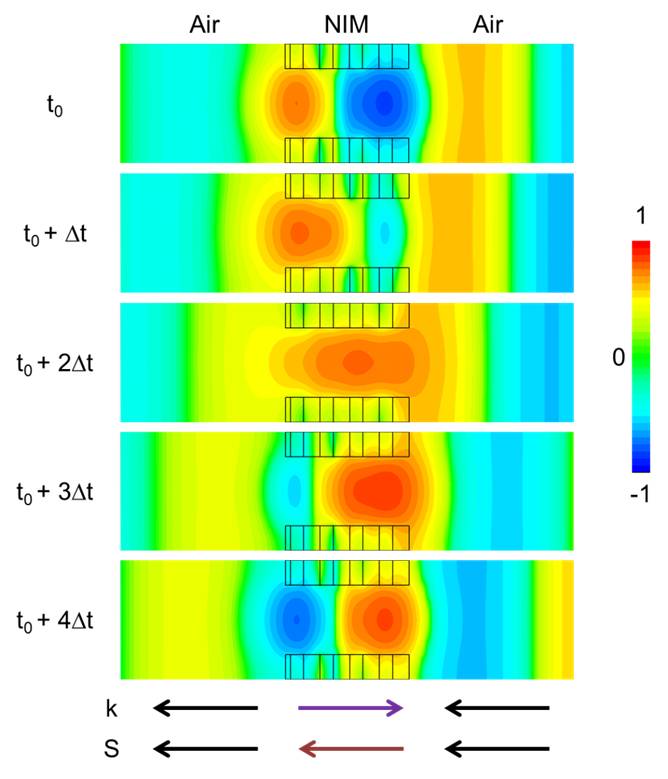}
\caption{\footnotesize \textbf{Time evolution of the phase front at 790 nm}. While the Poynting vector S is always conserved, the wave vector k is shown to be anti-parallel inside the negative index metamaterial. Such backward propagating waves behavior is unique for a material with negative refractive index and negative phase accumulation.} \label{sample7m}
\end{center}
\end{figure}

To illustrate the negative phase delay of the fishnet negative index metamaterial (NIM), we show in Fig. \ref{sample7m} the time evolution of the propagating phase fronts inside the bulk metamaterial at 790 nm wavelength, with the cross section taken at the center of the fishnet holes. The color map represents the normalized electric field in the x-direction. S and k are the Poynting vector and the wave vector, respectively. Unlike conventional positive refractive index metamaterials, the Poynting vector and wave vector are essentially antiparallel inside the NIM, demonstrating the negative phase accumulation and backward wave propagation behavior. 

\noindent\textbf{Alignment of Metamaterial in Sagnac Interferometer ---}
In our experiment, the NIM is mounted on an automated translation stage so that it can be reliably and repeatably removed and inserted. 
It has a clear aperture of approximately $20$ $\mu$m, thus we focus the beam sufficiently to pass through it. 
To find the optimal position of the NIM, we scan the translation stage, while monitoring the transmission of both the clockwise and counter-clockwise modes of the Sagnac interferometer.
We align the sample, relative to the focus of the lenses, such that the transmission of both modes is maximised at the same position.Another point of concern is the significant back reflection ($\approx50\%$) of the NIM for our wavelength range. Since this back reflection can couple to our detectors, we slightly tilt the NIM, by $0.44^\circ$, to reduce this background signal. 
We tilt the NIM along an carefully chosen axis so as to keep the polarization parallel to the thinner lines of the fishnet nanostructures, it has be shown that in this configuration such metamaterials still work optimally \cite{minovich2010tilted}. 

\end{document}